\documentclass[a4paper,11pt]{article}
\usepackage{pos}
\usepackage[normalem]{ulem}

\title{Primordial Kerr Black Holes}
\author*[a,b,c]{Alexandre Arbey} 
\author[a]{J\'er\'emy Auffinger}
\author[d,e,f]{Joseph Silk}

\affiliation[a]{Universit\'e de Lyon, Universit\'e Claude Bernard Lyon 1, CNRS/IN2P3, 
Institut de Physique des 2 Infinis de Lyon, UMR 5822, F-69622, Villeurbanne, France}
\affiliation[b]{Institut Universitaire de France, 103 boulevard Saint-Michel, 75005 Paris, France}
\affiliation[c]{Theoretical Physics Department, CERN, CH-1211 Geneva 23, Switzerland}
\affiliation[d]{Sorbonne Universit\'e, CNRS, UMR 7095, Institut d'Astrophysique de Paris, 98 bis bd Arago, 75014 Paris, France}
\affiliation[e]{Department of Physics and Astronomy, Johns Hopkins University, Baltimore MD 2218, USA}
\affiliation[f]{Beecroft Institute of Particle Astrophysics and Cosmology, University of Oxford, Oxford OX14BN, UK\\~}

\emailAdd{alexandre.arbey@ens-lyon.fr}
\emailAdd{auffinger@ipnl.in2p3.fr}
\emailAdd{joseph.silk@physics.ox.ac.uk}

\abstract{
Primordial Black Holes (PBHs) are appealing candidates for dark matter in the universe but are severely constrained by theoretical and observational constraints. We will focus on the Hawking evaporation limits extended to Kerr black holes. In particular, we will discuss the possibility to distinguish between black holes of primordial and of stellar origins based on the Thorne limit on their spin. We will also review the isotropic extragalactic gamma ray background constraints and show that the ``window'' in which PBHs can constitute all of the dark matter depends strongly on the PBH spin. Finally, we will consider the possibility that the so-called Planet 9 is a primordial black hole.
}

\FullConference{%
  40th International Conference on High Energy physics - ICHEP2020\\
  July 28 - August 6, 2020\\
  Prague, Czech Republic (virtual meeting)
}

\begin{document}

\maketitle

\section{Introduction}

Black holes (BHs) are currently under scrutiny, and are particularly studied via the observations of gravitational waves (GWs) emitted during the mergers of black holes. Primordial black holes (PBHs) are specific cases of black holes which have been generated in the early Universe. They are known to be potential dark matter candidates, and span a broad range of masses~\cite{Carr:2009jm}. Assuming that the formation of PBHs occurred at the rate of one maximal mass BH per Hubble volume, the mass of PBHs formed at the cosmological time $t_0$ is given by
\begin{equation}
M_{\rm PBH} \sim M_{\rm Planck} \times \frac{t_0}{t_{\rm Planck}} \sim 10^{38}\;{\rm g}\;\times t_0( {\rm s})\,.
\end{equation}
In particular, PBHs formed at Planck time have the Planck mass, PBHs formed at $t_0\sim10^{-23}\,$s have a mass of $M\sim10^{15}\,$g, and PBHs formed at 1\,s have $M\sim10^5\,M_\odot$.

Predictions for the spins of PBHs are highly model-dependent, and there exist two classes of models mainly leading either to non-spinning PBHs (e.g. \cite{DeLuca:2019buf}) or to maximally spinning PBHs (for example \cite{Harada:2017fjm}). In the following, we will assume that PBHs can have any spin.

\section{Black hole spins}

Spinning black holes are described by the Kerr metric
\begin{equation}
 c^2 d\tau^2 = \big(c dt - a \sin^2\theta d\phi\big)^2\, \frac{\Delta}{\Sigma} - \left(\frac{dr^2}{\Delta} + d\theta^2\right) \Sigma - \big((r^2+a^2)d\phi-a \, c dt\big)^2 \, \frac{\sin^2\theta}{\Sigma} \,,
\end{equation}
where $M$ is the BH mass, $J$ its angular momentum, $a=J/Mc$, $\Sigma=r^2 + a^2 \cos^2\theta$, $\Delta = r^2 - R_s r + a^2$, $R_s=2 GM/c^2$. The horizon of Kerr BHs \sout{exists but} is deformed and flattened in comparison to Schwarzschild BHs. One generally refers to the reduced spin of BHs $a^*=J/M^2$ (in natural units where $c=\hbar=1$), and the maximal reduced spin corresponds to $a^*=1$.

\begin{figure}[!t]
\begin{center}
 \includegraphics[height=5.cm]{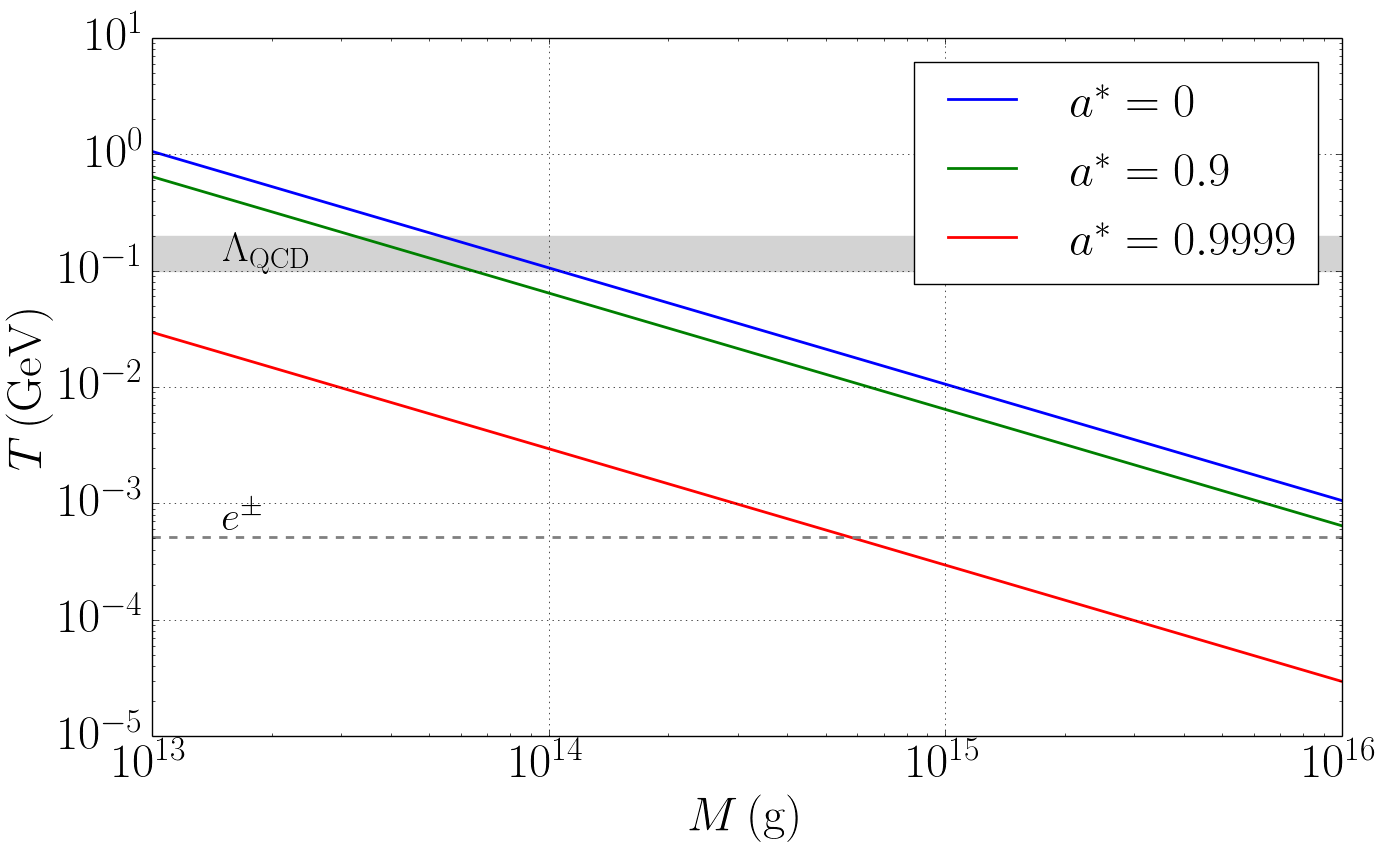}
\caption{Temperature of primordial black holes as a function of the black hole mass, for different values of the reduced spin $a^*$. Horizontal lines show limits on Hawking temperature from QCD considerations and from cosmic ray positrons. \label{fig:temperature}}
\end{center}
\end{figure}

\begin{figure}[!b]
\begin{center}
 \includegraphics[width=11.cm]{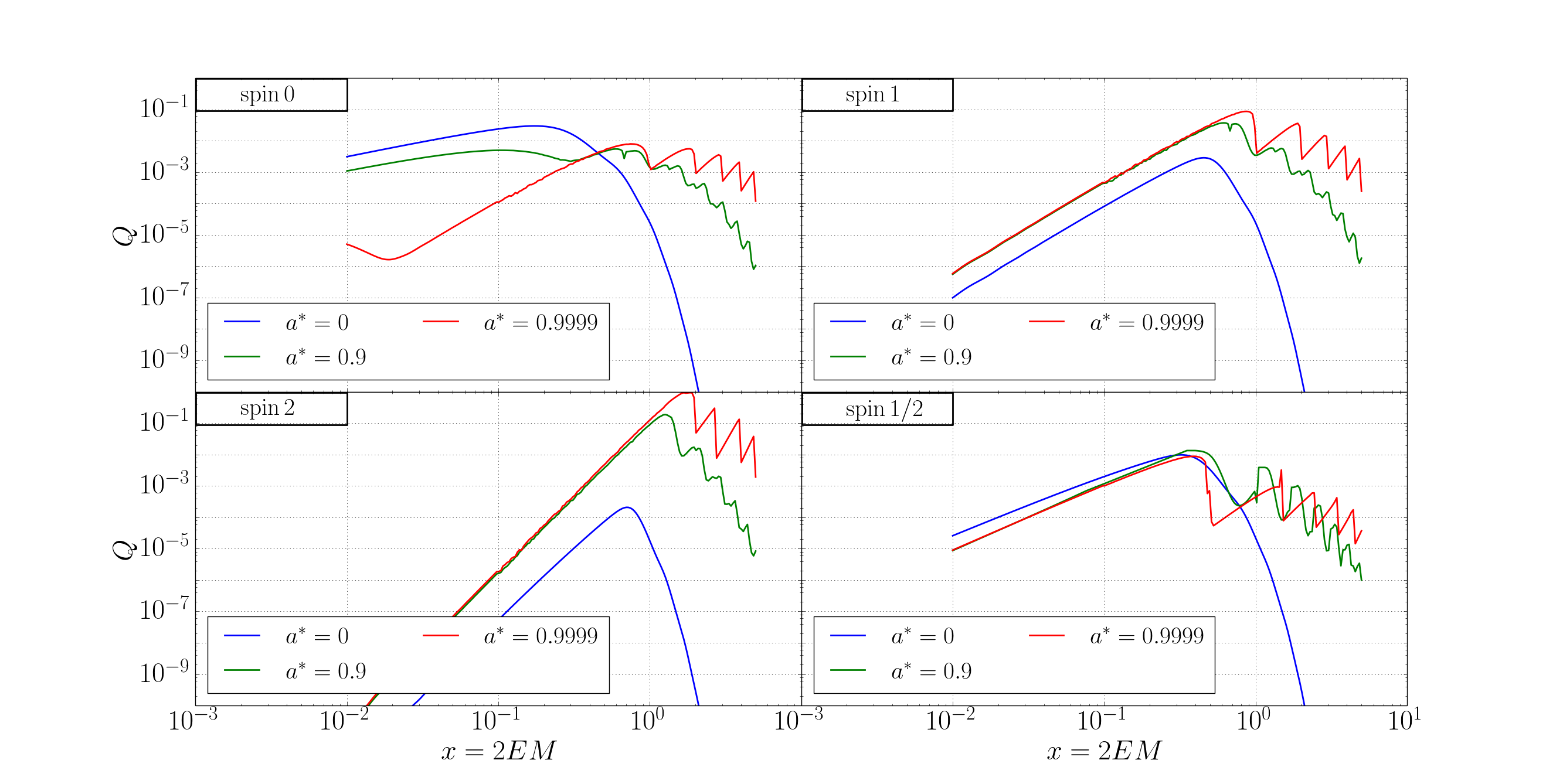}
\caption{Fluxes of emission $Q \equiv d^2N/dtdE$ (in GeV$^{-1}\cdot$s$^{-1}$ by a single black hole of spin-0, spin-1, spin-2 and spin-1/2 particles as a function of $2EM$, where $M$ is the black hole mass and $E$ the particle energy, for different reduced spins $a^*$.\label{fig:luminosities} } 
\end{center}
\end{figure}

Light BHs are known to evaporate by emitting Hawking radiation \cite{Hawking:1974sw}. In particular, a temperature can be associated to BHs, as seen in Figure \ref{fig:temperature}, and the emission spectrum of Hawking radiation is similar to the one of blackbody radiation. %
To make accurate predictions for Hawking radiation, we have written the code {\tt BlackHawk} \cite{Arbey:2019mbc}, which incorporates precise calculations of the so-called greybody factors and decay and hadronization of the \sout{produced} emitted particles, and which was used to produce the following figures. In particular, we precisely computed the emission spectra of particles of spins 0, 1, 2 and 1/2 by PBHs are shown in Figure~\ref{fig:luminosities} for different reduced spins. One can see that the spin can strongly affect the emission spectra.

\begin{figure}[!t]
\begin{center}
 \includegraphics[height=4.7cm]{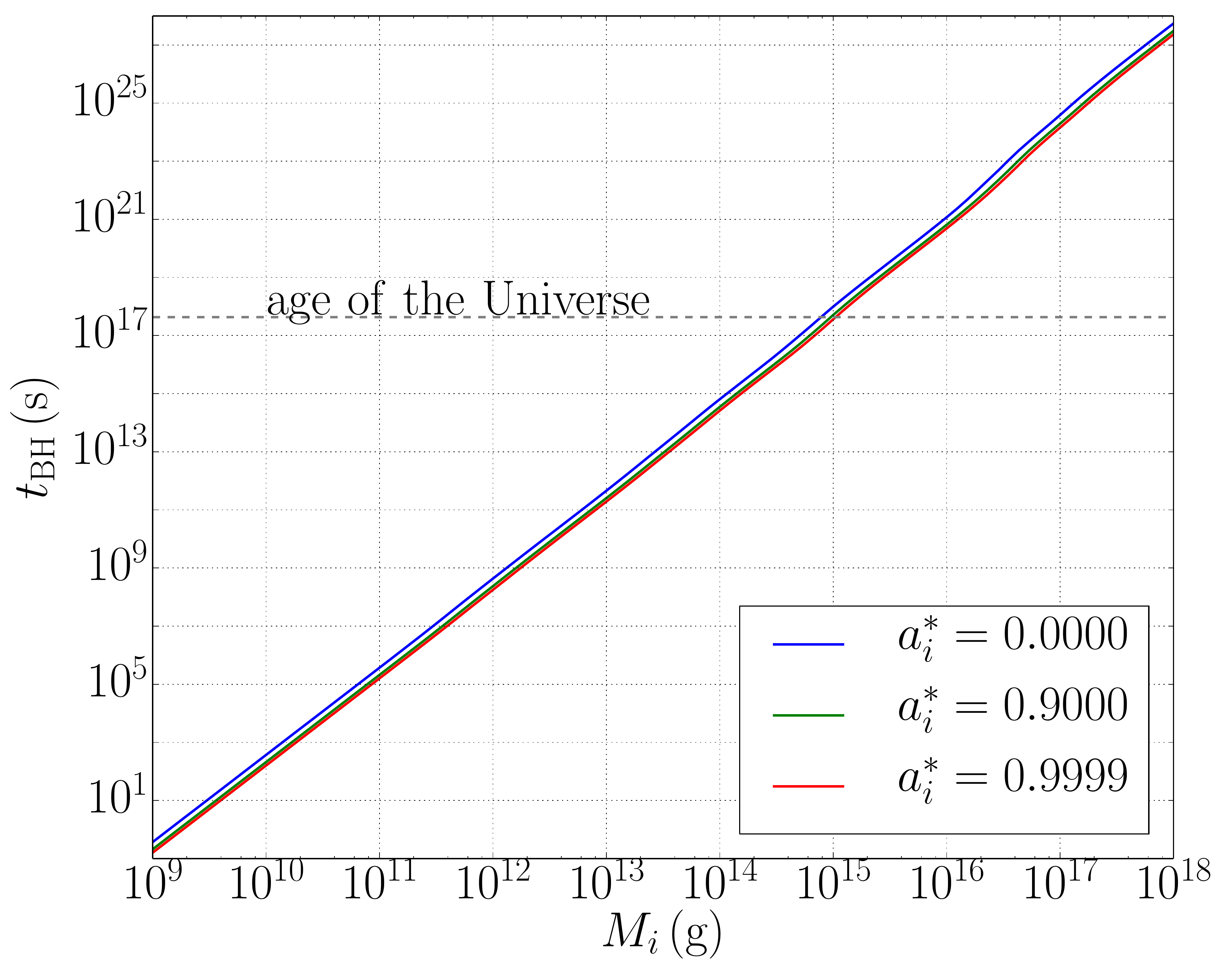}~~~\includegraphics[height=4.7cm]{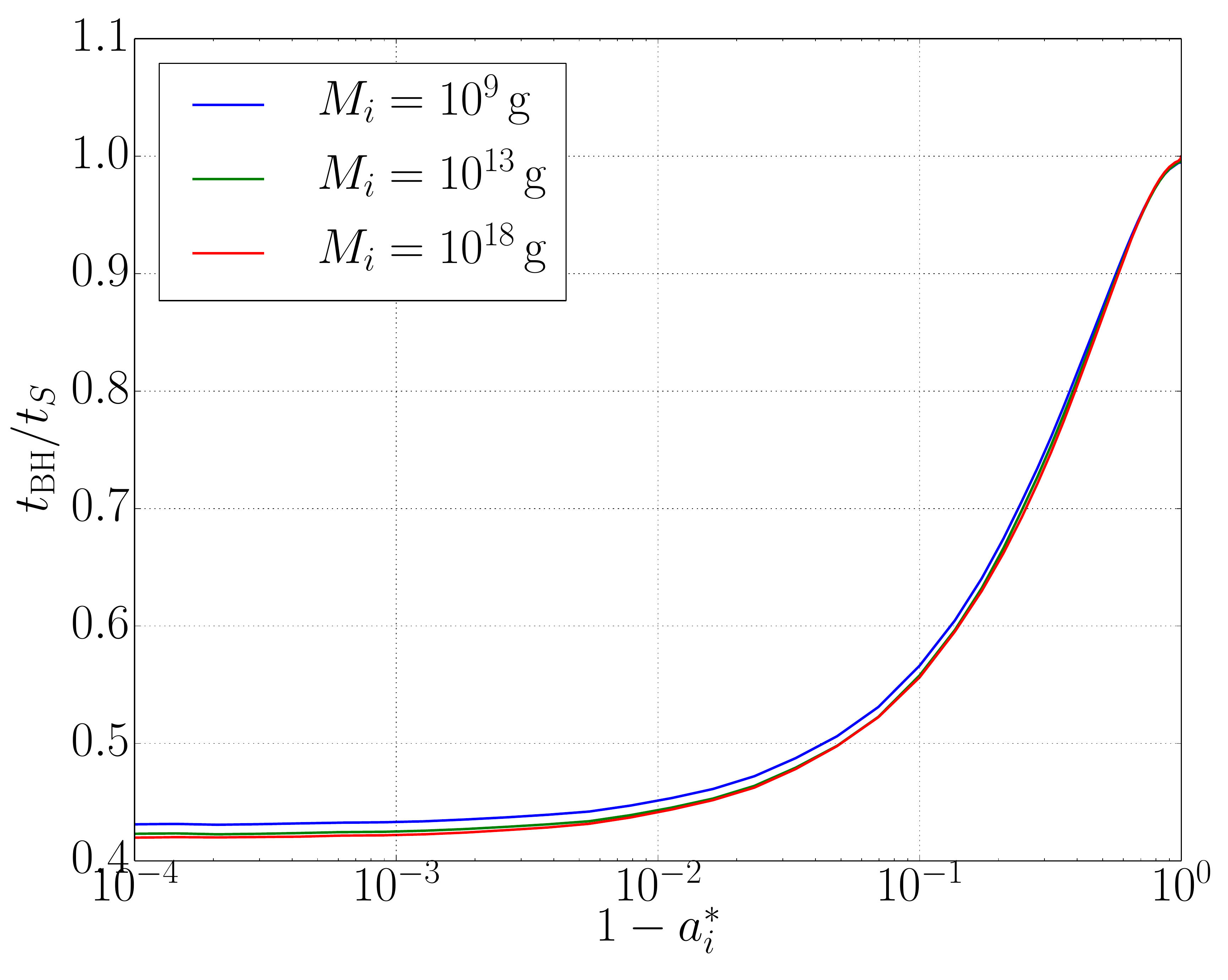}
\caption{Left: lifetime of primordial black holes as a function of their initial mass, for different values of the initial reduced spin $a_i^*$. Right: Lifetimes of $10^{10}$ and $10^{19}\,$g black holes as functions of $a_i^*$, relative to the lifetime of a non-spinning black hole of equal mass. (From \cite{Arbey:2019jmj}.)\label{fig:lifetime}}
\end{center}
\end{figure}

The emission of Hawking radiation results in a decrease of the \sout{energy} mass of BHs, which undergo a vanishing process. The lifetime of BHs is shown in Figure~\ref{fig:lifetime}, where one can see that PBHs with masses below $10^{15}$ g have already vanished since their production in the early Universe, and that the existence of a spin does not strongly affect the lifetime.

\begin{figure}[!b]
\begin{center}
 \includegraphics[height=4.7cm]{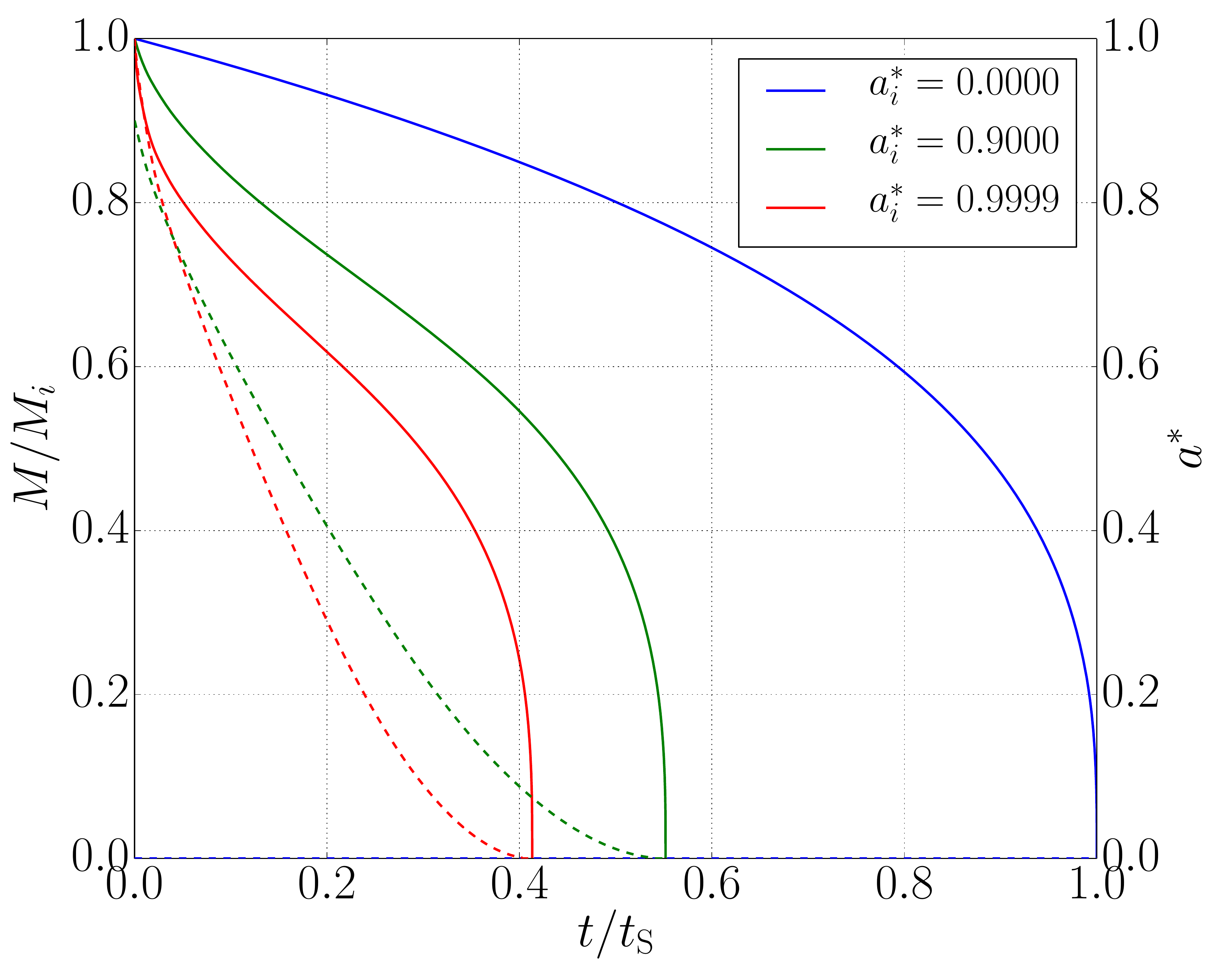}~~~~~\includegraphics[height=4.7cm]{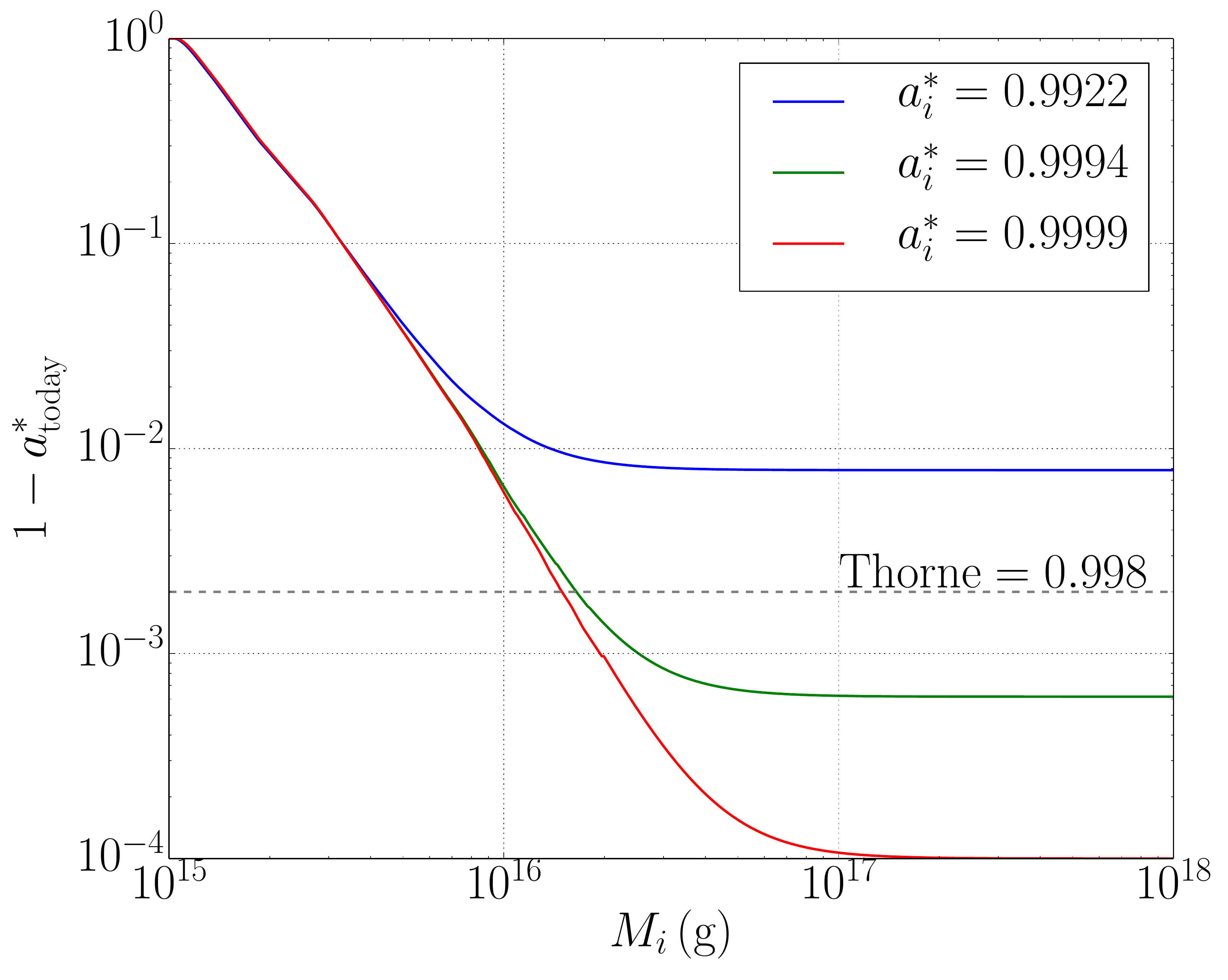}
\caption{Left: evolution of the mass (solid line) and spin (dotted line) as functions of time, for different initial reduced spin $a_i^*$. Right: Current values of the reduced spins of black holes with reduced initial spin 0, 0.99 and 0.9999, as functions of the initial black hole mass. The Thorne line delimits the maximal spin of black holes of stellar origin. (From \cite{Arbey:2019jmj}.)\label{fig:evolution}}
 \end{center}
\end{figure}

In Figure~\ref{fig:evolution}, we show the evolutions of the mass and spin of Kerr BHs with time for different values of their initial spin. We also compute the spin that PBHs can reach today depending on their initial mass and spin, and conclude that PBHs with initial reduced spins above 0.999 can still have today a spin above the Thorne limit, as long as their inital mass is above $2\times 10^{16}\,$g. This is an important result since the Thorne limit corresponds to the maximal value that stellar BHs are expected to have \cite{Thorne:1974ve}, and therefore the detection of many BHs with spins above the Thorne limit would be an indication towards a primordial origin. 

\section{Detectability of Hawking radiation}

We now consider the possibility to detect PBHs via their Hawking radiation. For this, we assess the impact of Hawking radiation of PBHs on the extragalactic gamma-ray background (EGRB) radiation, and derive constraints on the PBHs from observational data. A detailed description of the analysis can be found in Ref.~\cite{Arbey:2019vqx}.

\begin{figure}[!t]
\begin{center}
 \includegraphics[width=10.cm]{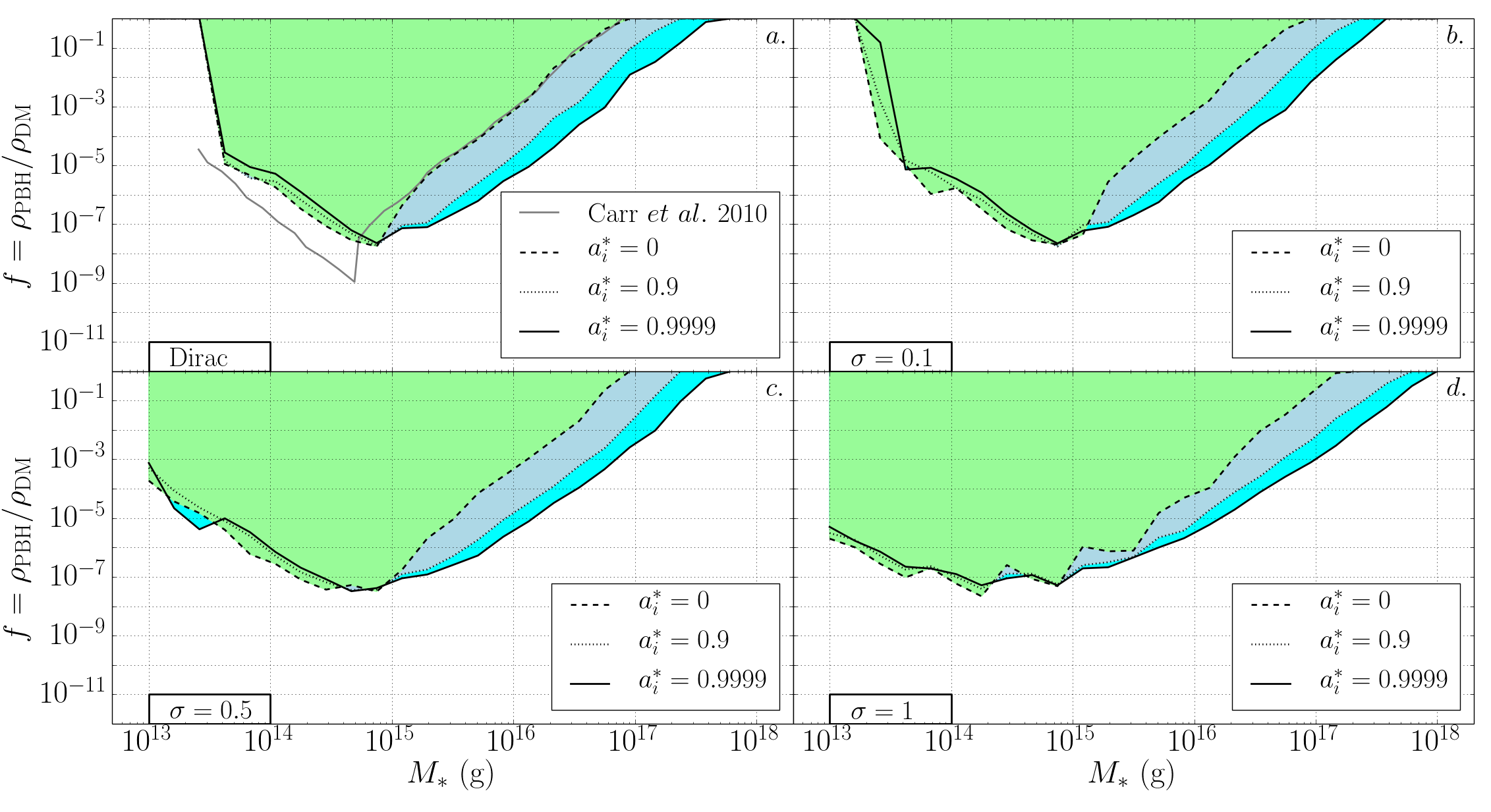}
\caption{In the plane of the fraction of PBHs relative to dark matter vs. PBH mass, regions excluded by EGRB radiation observations, for different values of the initial spin, and for different distributions of the PBH masses. The grey line corresponds to the exclusion limit derived in Ref.~\cite{Carr:2009jm}. (From \cite{Arbey:2019vqx}.)\label{fig:egrb}}
\end{center}
\end{figure}

In Figure~\ref{fig:egrb}, we show the constraints on the fraction of the PBH density relative to the dark matter density as a function of the PBH mass, for different initial spins. We consider different distributions for the PBH masses: monochromatic spectrum (Dirac distribution), and log-normal distribution with variance $\sigma = 0.1$, 0.5 and 1. The results show that the strongest constraint is obtained for masses about $10^{15}$\,g, and can extend up to $10^{18}$\,g. The initial distribution affects the results, as well as the initial spin of PBHs.

\begin{figure}[!b]
\begin{center}
 \includegraphics[width=8.cm]{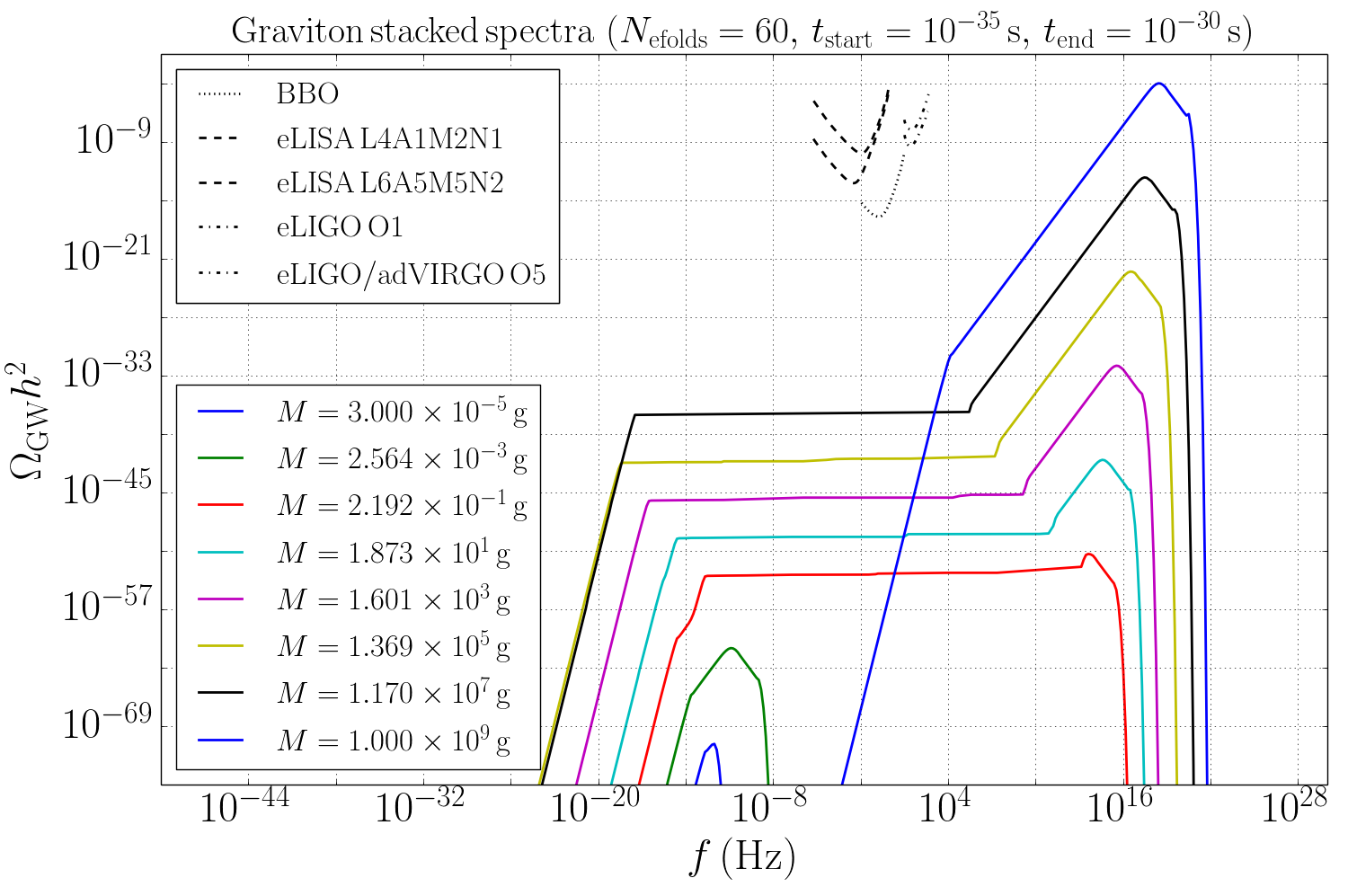}
\caption{Density of gravitational waves emitted by evaporated PBHs at the beginning of the Universe as a function of their frequency, for different initial masses of PBHs. We assume a cosmological scenario with an inflation period of 60 e-folds. The limits of BBO, eLISA and eLIGO/adVirgo are superimposed for comparison. \label{fig:gw}}
\end{center}
\end{figure}

It is interesting to notice that if gravity is mediated by a graviton particle, Hawking radiation \sout{can contain} applies to gravitons, which results in the fact that PBHs emit gravitons, which can be interpreted in terms of GWs. The background of GWs emitted by PBHs which are still present today is however too weak to be observed, even with the future generations of GW detectors. However, contrary to radiation, GWs emitted at the beginning of the Universe can still be observable today. In Figure~\ref{fig:gw} we show the density of GWs emitted at the beginning of the Universe by PBHs which have vanished today, as a function of their frequency, for different initial masses of PBHs. For PBHs of $10^9$\,g, the GW density can reach the sensitivity of the current GW experiments, but with frequencies around $10^{16}$\,Hz, which would necessitate extremely small and precise GW detectors.

\section{P9, a black hole in the Solar System?}

Concordant observations seem to point towards the existence of a ninth planet in the Solar System, at more than 400 AU from the Sun. However, since no planet has been observed with conventional telescopes, it has been proposed in Ref.~\cite{Scholtz:2019csj} that it could instead be a PBH, named P9.  

Should this hypothesis be true, such a PBH in the Solar System would constitute a wonderful laboratory for  studying black hole properties and quantum gravity effects. We have shown in Ref.~\cite{Arbey:2020urq} that it would be impossible to detect Hawking radiation of P9 from Earth, and not even with a probe travelling towards P9, because of the blackbody radiation from the Cosmic Microwave Background (CMB). However, a satellite  probe orbiting  P9 would allow the  possibility  of detection and study of Hawking radiation, since the BH horizon would screen radiation from the CMB.

\begin{figure}[!t]
 \includegraphics[height=4.5cm]{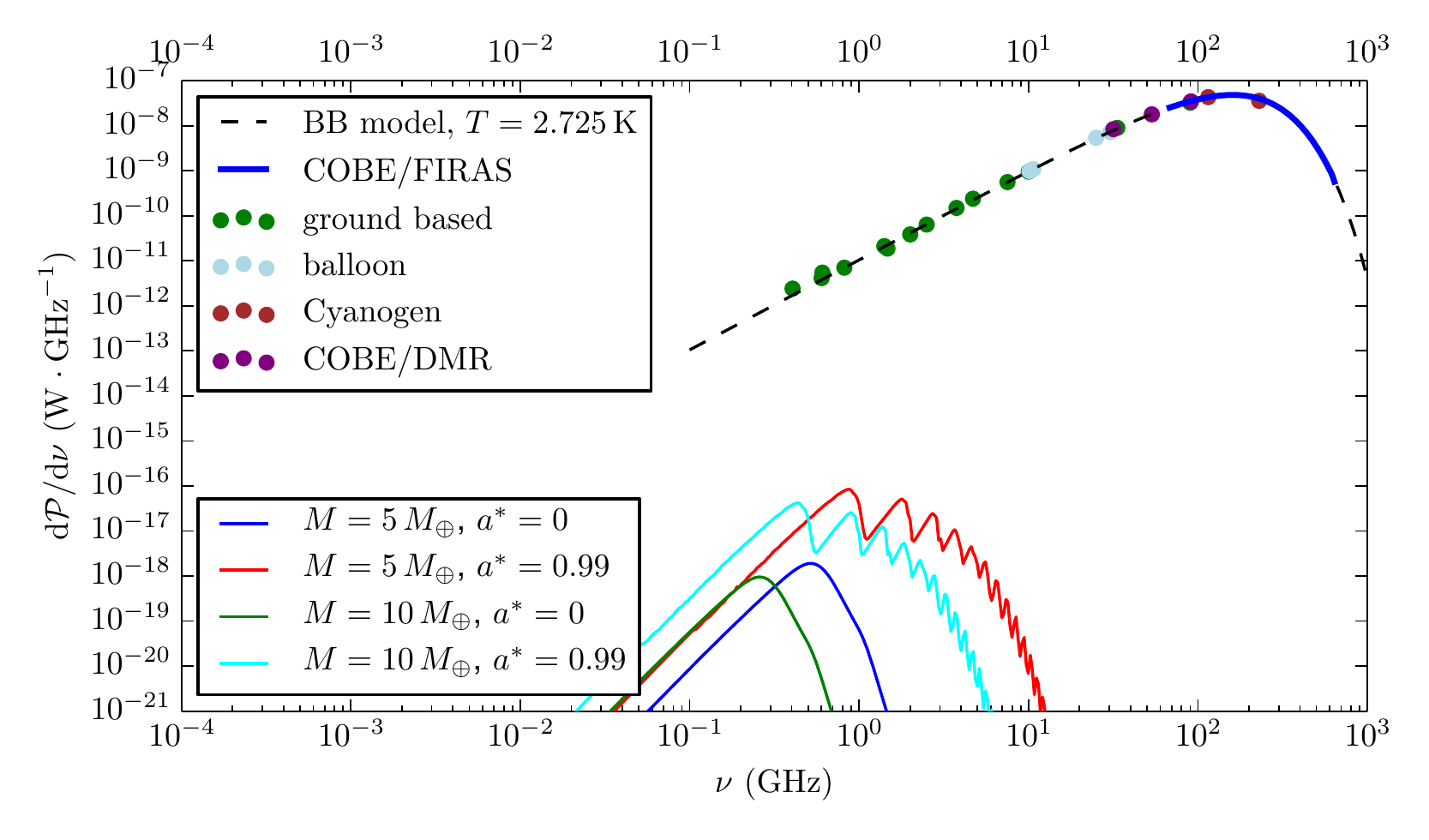}\includegraphics[height=4.5cm]{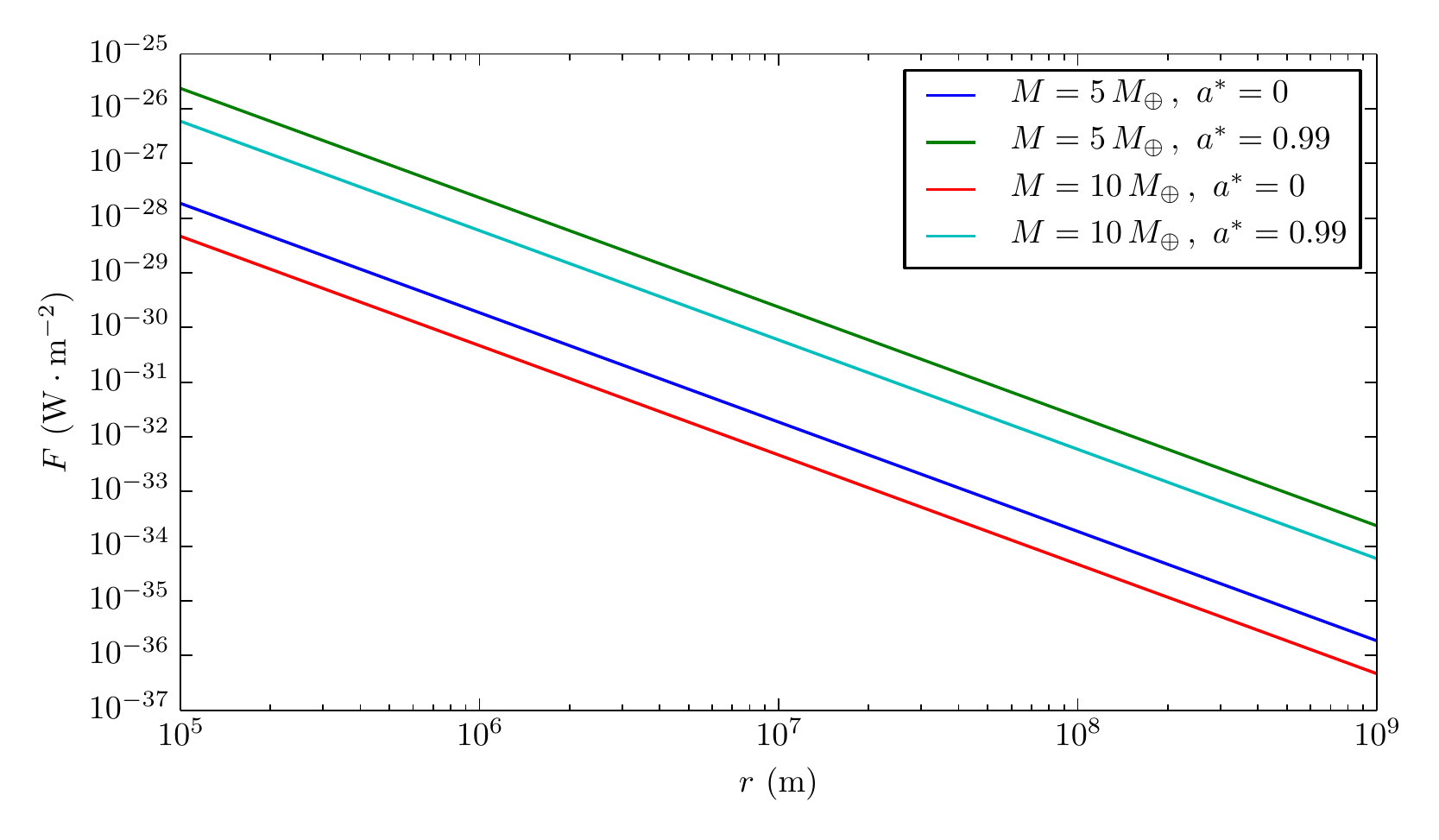}
\caption{Left: Hawking radiation spectrum of P9 as a function of the frequency. The CMB emission spectrum is shown for comparison. Right: Flux received by a satellite as a function of the orbital radius.\label{fig:P9}}
\end{figure}

In Figure~\ref{fig:P9} we show the Hawking radiation spectrum for photons, for different values of the mass and spin of P9. This reveals that the maximum emissivity is close to the GHz, making possible the use the precision antenna technology developed for wireless internet on Earth. We also plot the electromagnetic flux received by a satellite from the Hawking radiation of P9, as a function of the distance from P9.

\section{Summary}
Primordial black holes are potential candidates for dark matter. Scenarios of formation do not impose strong constraints on the spins of PBHs, which are therefore rather unconstrained. We have discussed the possibility of having nearly extremal spin PBHs and studied constraints from EGRB radiation. We have also considered P9 as a potential PBH in the Solar System.  This could therefore constitute an ideal lab to study a PBH.

\bibliographystyle{h-physrev5}
\bibliography{biblio}

\end{document}